\documentclass[epj,final]{svjour}
\usepackage{graphicx}
\usepackage{epsfig}
\usepackage{amsmath,amssymb}
\usepackage{color}

\begin{document}

\authorrunning{J. Erler et al}
\title{A stabilized pairing functional}
\author{J. Erler\inst{1} and P. Kl\"upfel\inst{1}  \and P.--G. Reinhard\inst{1}
} 
\institute{ 
Institut f{\"u}r Theoretische Physik, Universit{\"a}t Erlangen,
Staudtstrasse 7, D-91058 Erlangen, Germany
}
\date{\today / Received: date / Revised version: date}
\abstract{ 
We propose a modified pairing functional for nuclear structure
calculations which avoids the abrupt phase transition between pairing
and non-pairing states. The intended application is the description of
nuclear collective motion where the smoothing of the transition is
compulsory to remove singularities. The stabilized pairing functional
allows a thoroughly variational formulation, unlike the Lipkin-Nogami
(LN) scheme which is often used for the purpose of smoothing.  First
applications to nuclear ground states and collective excitations
prove the reliability and efficiency of the proposed stabilized
pairing.
} 

\maketitle

\section{Introduction}

Pairing is a key ingredient in a mean field description of nuclear
structure \cite{Boh58a}. The actual handling has much evolved. Early
models referred to simple forms assuming a constant gap or constant
pairing matrix element \cite{Rin80aB}. Modern nuclear mean field
calculations use a fully fledged pairing force, often even density
dependent, for an up to date review see \cite{Ben03aR}.  In any case,
the theoretical description relies on an extended notion of a mean
field, now including the two-quasi-particle density and gap-potential
also at mean field level. That allows to include a crucial part of
two-particle correlations at low expense. The price to be paid for
that simplification is the occurrence of a pairing phase
transition. Whenever the pairing strength crosses a critical value,
the system jumps from a pairing to a non-pairing state or vice versa.
For finite system such a sudden transition is an artefact
\cite{Gro97B}. A fully correlated treatment would produce correctly
smoothed transitions. That goal of smoothing can already be achieved
when working with a particle-number projected pairing state,
provided the mean-field variation is done after projection
(VAP). For then the system takes advantage of the pairing channel to
incorporate some correlations ``knowing'' that the exact particle
number will be restored. But the VAP method is cumbersome and not
applicable in connection with energy functionals \cite{Dob08a}. Thus
one employs usually simpler approaches which then are plagued by phase
transitions.
These are bearable when tracking ground states of nuclei where the
change in nucleon number is discontinuous anyway. However, they turn
into a severe problem when going along continuous changes as one does
for large amplitude collective motion, see
e.g. \cite{Ben03aR,Bra72aR,Rei87aR}. A sudden phase transition along
the collective evolution leads to unwanted singularities in the
description.  Within the adiabatic treatment of large-amplitude
collective motion, one encounters, e.g., singularities in the
collective masses.  A phase transition would be even more disastrous
for fully time-dependent mean-field calculations which enjoy a revival
those days, see e.g. \cite{Sim03a,Guo07a}. Thus there is a need to
smoothen the sudden pairing phase transition. A widely used recipe is
the Lipkin-Nogami (LN) scheme \cite{Lip60a,Nog64a} which aims at an
approximate treatment of particle-number projection and this
approximation has the similar smoothing effects as the exact
projection in VAP. Thus most of the
microscopic calculations of large-amplitude collective motion employ
LN pairing to generate the underlying series of collectively deformed
states, see e.g. \cite{Hee98b,Val00b,Fle04a,Ben06a}. However, the LN
prescription has still a few disadvantages in these large-scale
applications: it introduces complicated terms and one more equation
into the pairing scheme; the necessary feed-back on the
self-consistent mean field \cite{Rei96a} slows down the computational
speed and degrades the numerical stability; it is not strictly
variational which makes the validity of the LN approach less
controllable. Moreover, the feature of approximate particle-number
projection is not really exploited.  If a precise particle number is a
matter of concern, it is restored explicitely after the mean-field
calculations.  The superposition of mean-field states to a collective,
or correlated, state requires anyway an explicit handling of particle
number which is performed either by projection (e.g. \cite{Ben06a}) or
by adjustment in the average (e.g. \cite{Fle04a}).
As mentioned above, the formally perfect alternative of
particle-number projection with subsequent mean-field variation (VAP)
is inhibited in connection with the Skyrme-Hartree-Fock functional
\cite{Dob08a}.
There is thus a need for a conceptually simple, variationally
consistent, and numerically stable recipe to avoid, or smoothen, the
pairing phase transition, particularly for the simulation of
large-amplitude collective motion.  It is the aim of this paper to
present and to discuss a stabilized pairing functional which avoids
the phase transition while inducing a minimum of changes to other
nuclear properties.  The idea is to modify directly the pairing-energy
functional such that a vanishing pairing gap becomes associated with
infinitely increasing energy while merging into the ordinary
pairing-energy functional in regions of well developed pairing gap
such that standard pairing properties are not affected.  The
stabilized scheme is illustrated on a schematic two-level model.
First realistic application to nuclear structure and collective
excitations computed with the self-consistent Skyrme-Hartree-Fock
(SHF) method will also be presented.

\section{The stabilized functional}
\label{sec:model}

The standard procedure starts from an energy functional,
$E=E_\mathrm{mf}+E_\mathrm{pair}$, consisting out of a mean-field part
and an pairing functional. The energy depends on a set of
single-particle wavefunctions $\varphi_k$ and occupation amplitudes
$v_k$ with complementing non-occupation amplitudes
$u_k=\sqrt{1-v_k^2}$. Variation with respect to $\varphi_k^*$ yields
the mean-field equations and variation with respect to $v_k$ the gap
equation determining the pairing properties \cite{Rin80aB,Rei97c}.
The gap equation does not always have finite solution. There can
emerge a phase transition to zero pairing gap and zero pairing energy.
This induces singularities in the derivatives which spoils many
applications as, e.g., the description of large-amplitude collective
motion. 
The Lipkin-Nogami scheme (LN) avoids that sudden change by an
additional term $\propto\langle\hat{N}^2\rangle$ in the energy,
whereby $\hat{N}$ is the operator for particle number (for a discussion
in connection with self-consistent mean-field models see
\cite{Rei96a}). However, it has some disadvantages as discussed in the
introduction. 
We propose to stabilize a finite pairing by introducing a
counteracting term in the pairing energy functional. The stabilized
pairing functional reads
\begin{equation}
  E_{\rm pair}^{\mathrm{(stab)}}
  =
  E_{\rm pair}^{\mbox{}} 
  \bigg(1-\frac {E_{\rm cutp}^{2}}{E_{\rm pair}^{2}}\bigg)
  =
  E_{\rm pair}^{\mbox{}} 
  -
  \frac {E_{\rm cutp}^{2}}{E_{\rm pair}}
  \quad.
\label{eq:stabfunc}
\end{equation}
It can be applied to any given pairing functional $E_{\rm
pair}^{\mbox{}}$. It guarantees always a finite pairing because the
term $\propto E_{\rm pair}^{-1}$ grows huge when pairing is on the way
to breakdown. The stabilized function (\ref{eq:stabfunc})
introduces a new parameter, the cutoff pairing energy $E_{\rm cutp}$
which has to be chosen to deliver sufficient stabilization with
minimal effects in the well pairing regime.
The stabilized equations are derived as before by variation
of the total energy, now 
\begin{subequations}
\label{eq:stabeqs}
\begin{equation}
  \delta E
  =
  \delta\left[
   E_\mathrm{mf}^{\mbox{}}+E_\mathrm{pair}^{\mathrm{(stab)}}
  \right]
  =
  0
  \quad.
\end{equation}  
This yields the stabilized coupled mean-field and gap equations 
\begin{eqnarray}
  \hat h \psi_k  
  &=&
  \epsilon_k \psi_k 
  \quad,
\\
  2(\epsilon_{k}-\lambda) u_k v_k
  &=&
  \Delta_{k}^{\mathrm{(stab)}}(u^2_k-v^2_k)
  \quad,
\\
  \Delta_{k}^{\mathrm{(stab)}}
  &=&
  \Delta_{k}\bigg(1+\frac{E_{\rm cutp}^{2}}{E_{\rm pair}^{2}}\bigg)
 \quad,
\label{eq:stabgap}
\end{eqnarray}
\end{subequations}
where $\Delta_k$ is composed from the $u_k,v_k$ as in regular BCS
approach for the given functional. The detailed form depends on the
actual system. It will be specified later on in connection with the
examples.

\section{Test in a two-level model}

For a first exploration we employ a simple-most model for mean field
plus pairing. We consider $N$ particles in two energy levels, both $N$-fold
degenerated. The levels are labeled by $(s,m)$ where
$\mathrm{s}\in\{\mathrm{u},\mathrm{l}\}$ stands for the
principle quantum number and $m=\pm 1,...,\pm N/2$ accounts for the
(degenerated) magnetic quantum numbers.
The energy difference between upper and lower states,
$E_\mathrm{ul}=\varepsilon_\mathrm{u}-\varepsilon_\mathrm{l}$
simulates the shell spacing. Pairing is added with a constant pairing
force \cite{Rin80aB}. The occupation amplitudes are also degenerated
$v_{\mathrm{s}m}=v_\mathrm{s}$.  
For symmetry reasons, we have $v_\mathrm{l}=v$,
$v_\mathrm{u}=\sqrt{1-v^2}$, and $u=\sqrt{1-v^2}$.
This makes altogether the total energy
\begin{subequations}
\begin{eqnarray}
  E
  &=&
  E_\mathrm{mf}+E_\mathrm{pair}
  \quad,\quad
\\
  E_\mathrm{mf}
  &=&
  \frac{E_\mathrm{ul}}{2}N(u^2-v^2)
  \quad,\quad
\\
  E_\mathrm{pair}
  &=&
  -
  G(Nuv)^2
  \quad.
\end{eqnarray}
\end{subequations}
The mean-field variation is obsolete because the single particle
energies $\varepsilon_\mathrm{u,l}$ are given. It remains the
variation of pairing properties which yields the gap equation
$E_\mathrm{ul}uv+\Delta(u^2-v^2)=0$ where $\Delta=GNuv$.
Its solution  for the gap  becomes
\begin{equation}
  \Delta
  =
  \frac{E_\mathrm{ul}}{2}\sqrt{\left(\frac{NG}{E_\mathrm{ul}}\right)^2-1}
  \quad.
\end{equation}
The decisive parameter for the model is the ratio
$NG/E_\mathrm{ul}$. Pairing breaks down for $NG/E_\mathrm{ul}\leq
1$. The breakdown should now be hindered by introducing the stabilized
functional (\ref{eq:stabfunc}) which yields the stabilized gap as
\begin{equation}
  \Delta^{\mathrm{(stab)}}
  =
  \frac{E_\mathrm{ul}}{2}
   \sqrt{\left(\frac{NG}{E_\mathrm{ul}}\right)^2
         \bigg(1+\frac{E_{\rm cutp}^{2}}{E_{\rm pair}^{2}}\bigg)^2  
        -1}
  \quad.
\end{equation}
This is a rather involved equation for the gap because $E_{\rm pair}$
on the right hand side depends implicitly on $\Delta$. But the
crucial result is that the expression under the square root can never
vanish and thus the gap equation has always a finite real solution.

\begin{figure}
\centerline{\includegraphics[width=7.9cm]{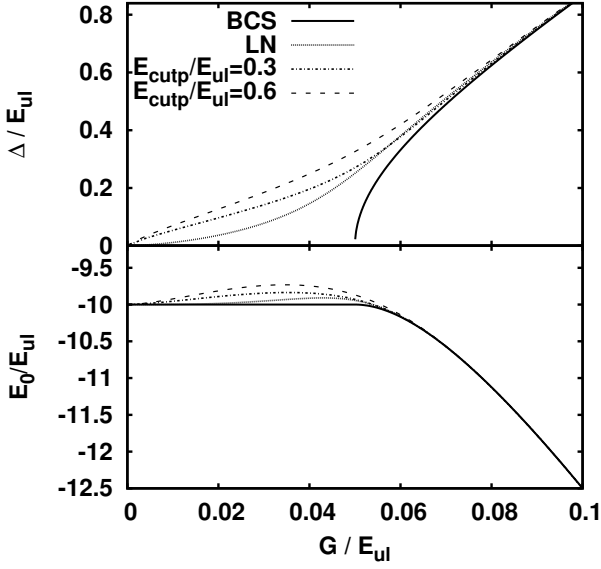}}
\begin{center}
\end{center}
\caption{\label{fig:twolevelmodel}
Upper: Dependence of
the pairing gap on the coupling strength for
various methods in comparison, BCS, LN and the
stabilized pairing with two different cutoff 
parameters as indicated.
Lower: The total energy of the system as function 
of the coupling strength.
}
\end{figure}
Figure \ref{fig:twolevelmodel} shows results of the two-level model
for phase space $N=20$. The pairing gaps are given in the upper
panel. BCS produces a phase transition at relative coupling strength
$G/E_\mathrm{ul}\approx 0.05$ and LN clearly removes that
discontinuity producing a gap which smoothly approaches zero in the
limit $G\longrightarrow 0$.
In the regime of active pairing, the LN gap approaches the BCS gap as
it should. Both results from the stabilized functional
(\ref{eq:stabfunc}) smoothen the transition, at first glance similar
to LN. However, the functional behavior differs in detail. The trend
at small $G$ produces more efficient
stabilization of the pairing gap, as can be seen, e.g., from the result for
the extremely small $G/E_\mathrm{ul}=0.01$. 
At the side of active pairing, the stabilized functional
also converges to the BCS curve where the speed of convergence
improves with decreasing $E_\mathrm{cutp}$. The goal is to have a
slow decrease of $\Delta$ for $G\longrightarrow 0$ combined with a
fast convergence to BCS for large $G$. The stabilized pairing
promises to provide here a better compromise than LN.
The lower panel of figure \ref{fig:twolevelmodel} shows the binding
energies. LN follows the BCS curve with a maximum deviation of 2\%
near the critical value $G/E_\mathrm{ul}\approx 0.05$. The stabilized
pairing shows also deviation which, however, extend further below the
critical value. The maximum deviation depends sensitively on the
cutoff energy. The smaller cutoff produces an error similar to LN and the
larger cutoff has a larger error. Thus a good compromise will dwell
at the lower side of cutoff energies. The actual choice will depend,
of course, on the actual system and the intended applications.

\begin{figure}
\centerline{\includegraphics[width=7.9cm]{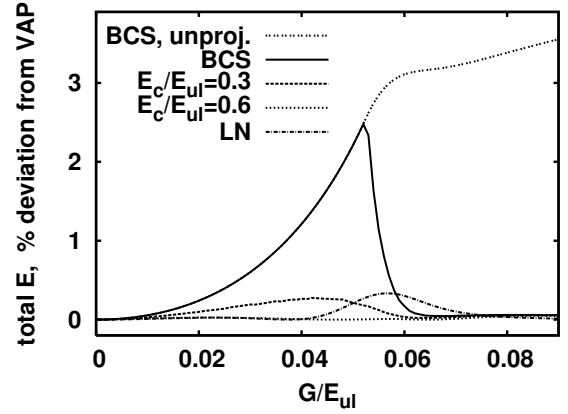}}
\begin{center}
\end{center}
\caption{\label{fig:PNP}
Comparison of various approaches with variation after particle-number
projection (VAP) for the total energy as observable drawn as
\% deviation from the VAP result versus the normalized coupling
strength. 
Test case is again a system with particle number $N=20$.
The four cases, BCS, LN and stabilized BCS at two different
cutoff energies employ  particle-number projected calculations 
applied to the given mean field state
(projection after variation = PAV). 
An unprojected mean-field result is also given 
for the case of BCS.
}
\end{figure}
The two-level model is based on an explicit Hamiltonian.  This allows
to perform particle-number projection.  The most advanced approach is
mean-field variation after projection (VAP). We compare that with the
results from mean-field calculations 
where projection is done after mean-field variation.
Figure \ref{fig:PNP} shows the results drawn as relative deviation
from VAP because the differences for the projected results are too
small to be seen when showing the total energies as such.  The BCS
case deviates most strongly below the critical coupling strength
$G$. Pairing was not active here and no energy can be gained from
pairing with subsequent projection. The deviation stays at the level
of 3--4 \% for pure mean-field results. The unprojected BCS curve
shows the typical size and trend which looks very similar also for the
(unprojected) other methods. The projected BCS result converges above
the critical strength rapidly to a very small deviation. LN and
stabilized gap show very small deviations throughout. The figure
confirms that LN with subsequenz projection is an efficient approach
to full VAP. And it shows that the newly developed stabilized gap
performs equally well, and sometimes better, in that perspective. The
case of $E_\mathrm{cutp}/E_\mathrm{ul}=0.6$ shows that proper tuning
may even enhance the quality. On the other hand, being satisfied with
LN quality means that one disposes of a broad band of choices for
$E_\mathrm{cutp}/E_\mathrm{ul}$ which one can use to optimize other
aspects as, e.g., stability.

\section{Nuclear ground states and collective dynamics}

\subsection{The modified functional and BCS equations}

Self-consistent nuclear mean-field models are based on effective
energy-density functionals. The three most widely used models are the
relativistic mean-field model (for reviews see e.g.
\cite{Ser86aR,Rei89aR,Rin96aR}), the Gogny force (see e.g.
\cite{Dec80a}), and the Skyrme-Hartree-Fock (SHF) approach (for a
recent review see e.g. \cite{Ben03aR}). We concentrate here on SHF.
Although there had been attempts to incorporate pairing directly into
the Skyrme force \cite{Dob84a}, most functionals keep mean-field and
pairing properties separate. Recent SHF calculations employ generally
a pairing functional derived from a zero-range pairing force
\cite{Ton79a,Kri90a}. It reads
\begin{subequations}
\begin{eqnarray}
  E_{\rm pair}
  &=&
  \frac{1}{4}\sum _\mathrm{q} 
  \int d^{3}r V_{\rm q}(\mathbf{r})\chi^{2}_{\rm q}(\mathbf{r})
  \quad,
\label{eq:E_pair}
\\
  \chi_{\rm q}(\mathbf{r})
  &=&
  2\sum_{k>0}\!f_k\, u_{k}v_{k}|\varphi_{k}(\mathbf{r})|^{2} 
  \quad,
\label{eq:chi}
\end{eqnarray}
where $f_k$ is a smooth cutoff function which limits the pairing space
to an energy region up to 5 MeV above the Fermi energy
\cite{Ben03aR,Ben00c}. That pairing functional is added to the SHF
mean-field energy functional which is rather lengthy. We refer to
\cite{Ben03aR,Sto07aR} for details.  
The state-dependent pairing gap from this functional becomes 
\begin{equation}
 \Delta_k
  = -
 \frac{V_q}{2}
  \int d^{3}r\,\chi_{\rm q(k)}(\mathbf{r})|\varphi_{k}(\mathbf{r})|^{2} 
  \quad.
\end{equation}
\end{subequations}
The gap equations derived from
that functional show also the possible breakdown of pairing. We
modify the functional (\ref{eq:E_pair}) to the stabilized pairing
(\ref{eq:stabfunc}). As a consequence, we deal now with the stabilized
equations (\ref{eq:stabeqs}). These will be solved in spherical
symmetry for the ground states of spherical nuclei \cite{Rei91b} and
in axial symmetry for computing the collective deformation path and
subsequent collective dynamics \cite{Fle04a}.
There is a broad variety of parameterizations of the SHF mean-field
functional \cite{Ben03aR}. The actual choice is not crucial for the
present tests of the stabilized pairing functional. We use here the 
parameterization SkI3 \cite{Rei95a}.

The pairing gap from the functional (\ref{eq:E_pair}) becomes state
dependent as $\Delta_k$. In order to  have one number characterizing
the gap, we define an average pairing gap as
\begin{equation}
\label{eq:spec_av_gap}
  \bar \Delta_q
  =
  \frac{\sum_{k\in q}f_ku_kv_k\Delta_k}{\sum_{k\in q}f_ku_kv_k}
  \, .
\end{equation}
The weight factor $u_kv_k$ concentrates averaging to a region near the
Fermi energy where pairing is most relevant. That definition is found
to be the most appropriate for comparisons \cite{Ben00c}.

\subsection{Results and discussion}

\begin{figure}
\centerline{\includegraphics[width=7.9cm]{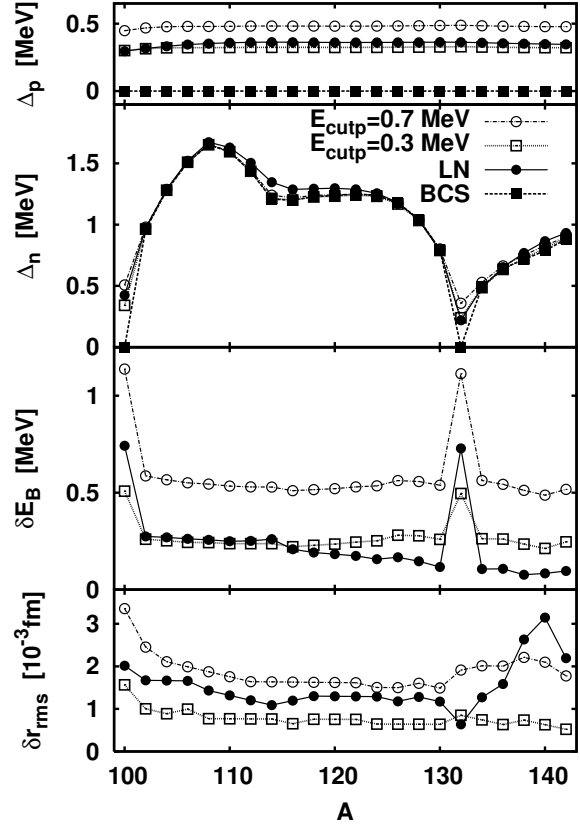}}
\caption{\label{fig:groundstates}
Ground state properties for the chain of
Sn isotopes computed with BCS, LN and stabilized
pairing at two $E_\mathrm{cutp}$.
Upper two panels:
Average proton and neutron pairing gaps.
Middle panel: difference $\delta E_\mathrm{B}$
of binding energy from
stabilized pairing or LN to the BCS binding energy.
Lowest panel: difference $\delta r_\mathrm{rms}$
for the charge r.m.s.radius.
}
\end{figure}
The uppermost two panels of figure \ref{fig:groundstates} show proton
and neutron pairing gaps along the semi-magic chain of Sn
isotopes.
The proton number Z=50 is magic throughout the isotopic chain.
Thus BCS proton pairing breaks down for all isotopes while LN and
stabilized pairing produce a small but finite gap. The actual size of the
stabilized gap depends somewhat on the chosen cutoff whereby the lower
$E_\mathrm{cutp}$ produces results very close to LN.
BCS neutron pairing drops to zero for the two magic neutron numbers
$N=50$ and $N=82$, has substantial pairing mid shell, and a short
transitional region near the magic numbers.  LN produces, of course,
a finite gap at the magic points and tries to come close to BCS when
going away from the shell closures. The results from stabilized
pairing do also produce finite gaps at the shell closures. The actual
value of the gaps there depend again somewhat on
$E_\mathrm{cutp}$. Outside the magic points, the gaps from stabilized
pairing come very close to the BCS values, for the small as well as
for the larger cutoff, and both closer than LN.

The lower two panels in figure \ref{fig:groundstates} show the effect on
bulk observables. Results are drawn as difference to the BCS
results in order to fit the small overall changes into a graphical
representation. The changes on the total binding energy (middle panel) have
an offset along all isotopes and a peak at the magic neutron
numbers. Offset and peak are expected because the proton
stabilization is always present and neutron stabilization adds to the
effect for N=50 and N=82.  The average size of the modification is
close to the LN values for the smaller cutoff. We have to mention,
however, that the energies from LN are an uncertain quantity because
the LN scheme is not variationally consistent. It is thus not fully
clear what really should be plotted here. The stabilized pairing, on
the other hand, has by construction a well defined energy. And a
change of the order of 0.2--0.5 MeV is well acceptable in view of a
typical uncertainty of 0.7--1 MeV for Skyrme forces.
The effect on radii (lowest panel) is extremely small as compared to
typical uncertainties of about 0.02--0.04 fm \cite{Ben03aR}. The same
holds for the whole density and deduced observables as diffraction
radius and surface thickness \cite{Fri82a,Fri86a}.

\begin{figure}
\centerline{\includegraphics[width=7.9cm]{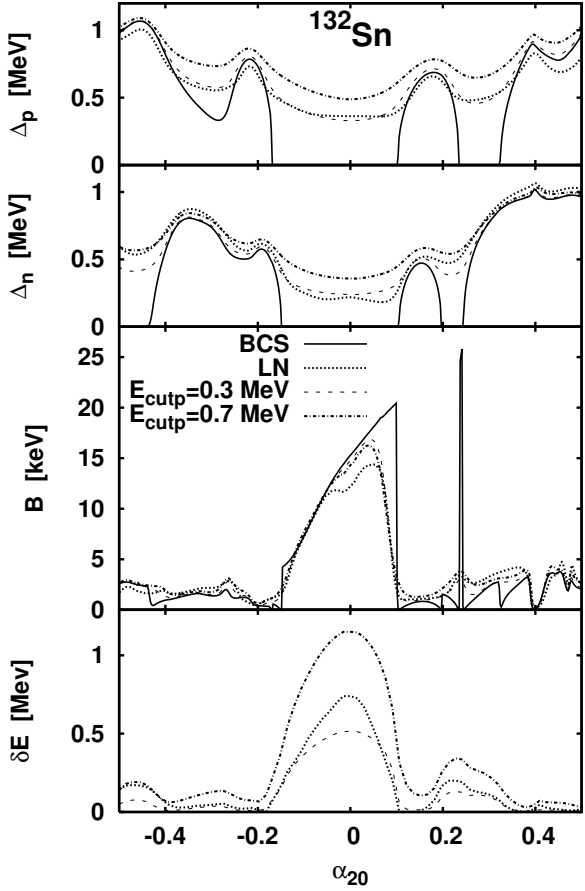}}
\caption{\label{fig:collective132Sn}
Proton and neutron gaps (upper two), collective mass for motion along $\alpha_{20}$ (middle), and 
difference of the collective potential from stabilized pairing or LN
to that from BCS
(lower) along the quadrupole deformation path for  $^{132}$Sn
computed with BCS, LN and stabilized pairing.
}
\end{figure}
Low lying nuclear $2^+$ states are related to large amplitude
collective motion. The nucleus vibrates along a collective deformation
path which consists out of a continuous series of quadrupole deformed
states. The path is generated by adding a quadrupole constraint to the
mean field equations sampling the various quadrupole
deformations; from the given path, one can derive then the ingredients
for the collective dynamics, the potential as expectation value of the
energy and the collective mass from double commutators with the
dynamical boost, for details see, e.g. \cite{Rei87aR,Fle04a}.
Figure \ref{fig:collective132Sn} shows results along the axially
symmetric quadrupole deformation path of $^{132}$Sn. BCS shows the
breakdown of neutron pairing near the spherical shapes and the mass
jumps immediately to large values because a pure Slater determinant is
more resistant to quadrupole motion than a well pairing BCS state.
The sudden changes make the BCS mass unusable for further processing
in the collective Hamiltonian.  LN and the stabilized functionals
succeed in preventing the breakdown of pairing and in producing a
smooth transition. They behave similar what the collective mass and
the changes in energy is concerned. It is to be noted that the
stabilized pairing remains closer to the BCS mass in the regime of
shell closure. The net effect of all these small differences on the
collective excitations will be an average over all deformations.

\begin{figure}
\centerline{\includegraphics[width=7.9cm]{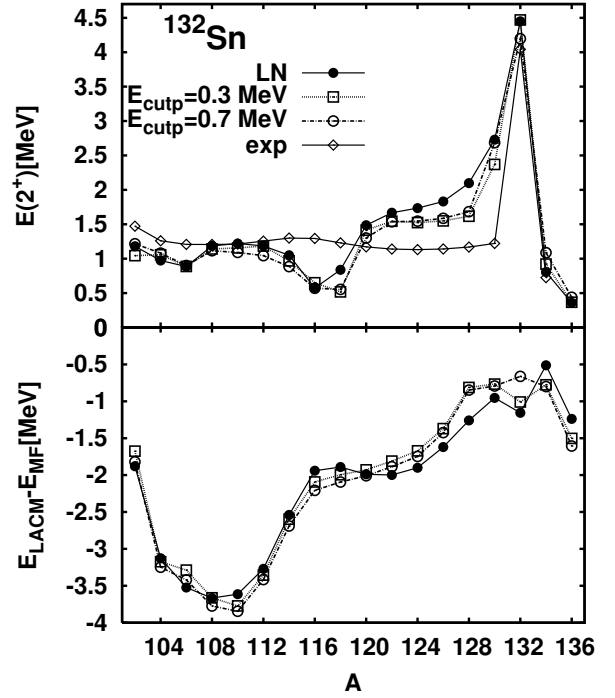}}
\caption{\label{fig:collspectra}
Systematic of quadrupole-collective properties
along the chain of
Sn isotopes computed with LN and stabilized pairing
at two $E_\mathrm{cutp}$.
Upper: The low-lying $E(2_1+)$ excitation energies.
Lower: Collective correlation energies.
}
\end{figure}
Figure \ref{fig:collective132Sn} has shown the impact of the
stabilized pairing functional and its $E_\mathrm{cutp}$ on the
constituents of the collective Hamiltonian, mass and potential. Figure
\ref{fig:collspectra} shows the net effect on the final states.
The results for LN and stabilized pairing are again similar in general
with a few interesting differences in detail. The sudden reduction of
the $E(2_1^+)$ when departing from the magic $N=82$ is very badly
reproduced by LN which produces a much smoothed transition. The
stabilized pairing performs better in that respect. The mid-shell
level is reached already at $N=78$ and $N=80$ also does a larger
down-step. The trend is the better reproduced the lower
$E_\mathrm{cutp}$. The excitation energies at the magic point and
mid-shell are similar for all methods which proves that the modified
pairing recipes do not affect too much the physics of the collective
excitations. The sometimes larger effects seen in the details of
collective mass and potential (figure \ref{fig:collective132Sn}) seem
to average out in the final collective energies.
The results for the correlation energies are even more robust.
The values differ in tiny detail, but that is not significant because the
correlation energy is anyway a small correction to the total energy.

The most important result concerns a practical aspect, not shown in
the above figures: Stabilized pairing is simpler to implement
numerically and it produces a faster and much more stable
iteration. This amounts to a factor three faster computation of the
collective deformation paths which is a substantial gain for large
scale applications.

\section{Conclusions}

We have investigated a new scheme to smoothen the pairing phase
transition in nuclei. The stabilized pairing is achieved by modifying
the pairing functional such that the pairing energy diverges for
vanishing gap. The modification is restricted to the regime of
small coupling and has little effect elsewise.
The intended effect should come out similar to what is usually
achieved by the Lipkin-Nogami (LN) scheme. 

The operation of the stabilized functional was demonstrated in a
schematic two-level model and compared to the LN. It was shown that a
smooth transition to vanishing gap can indeed be achieved by the
stabilized pairing functional while the convergence towards BCS results in
the well pairing regime is even somewhat faster for stabilized
pairing as compared to LN.
The BCS ground state with stabilized pairing can also serve
very well as starting point for subsequent particle number projection.
The results come very close to fully fledged mean-field variation
after projection.

The stabilized functional has been applied to a bunch of realistic
nuclear structure calculations, ground states of spherical nuclei and
low-energy collective dynamics.  All results show that stabilized
pairing is a valid alternative to LN. The effects are very similar in
general. There are some small differences in detail which give a
slight preference to stabilized pairing.

The results depend somewhat on the cutoff parameter inherent in the
model. Pairing properties are, of course, more sensitive, but bulk
properties, fortunately, not so much.  A good choice which has small
side-effecs and comes in many aspects close to LN is a small value
around $E_\mathrm{cutp}=0.3$MeV.

The major advantages of stabilized pairing lie in a formal and in a
numerical feature.  The formal advantage is a thoroughly variational
formulation. The modification is done only at the side of the
functional which provides a clean definition of the energy.  The
numerical gain is dramatic. Stabilized pairing provides factor three
faster convergence than with LN due to the simpler algorithm possible
(remaining practically identical to the robust BCS scheme). This is
the strongest argument in favor of the new scheme.
A stabilized pairing functional with fully variational formulation
will become inevitable in future time-dependent applications where a
possible phase transition would be disastrous and where LN is not
applicable.

\section*{acknowledgments}
That work was supported by the BMBF contract number 06 ER 124.  We
thank J. Maruhn (Frankfurt) for inspiring and instructive discussions.

\bibliographystyle{epj}
\bibliography{biblio,reviews,books,add}

%

\end{document}